\documentclass[twocolumn,trackchanges]{aastex63}
\hypersetup{linkcolor=red,citecolor=blue,filecolor=cyan,urlcolor=magenta}

\newcommand{\Msun}{$\mathit{M_{\odot}}$}
\newcommand{\Mch}{$\mathit{M_{\mathrm{Ch}}}\ $}
\newcommand{\Rnu}[1]{\uppercase\expandafter{\romannumeral{#1}}}

\shorttitle{The interaction of SN Iax ejecta with a He star}
\shortauthors{Y. Zeng, Z.-W. Liu \& Z. Han}

\begin{document}

\title{The interaction of Type Iax supernova ejecta with a helium companion star}

\correspondingauthor{Zheng-Wei Liu}
\email{zwliu@ynao.ac.cn}

\author{Yaotian Zeng}
\affiliation{Yunnan Observatories, Chinese Academy of Sciences (CAS), Kunming 650216, P.R. China}
\affiliation{Key Laboratory for the Structure and Evolution of Celestial Objects, CAS, Kunming 650216, P.R. China}
\affiliation{University of Chinese Academy of Science, Beijing 100012, P.R. China}

\author{Zheng-Wei Liu}
\affiliation{Yunnan Observatories, Chinese Academy of Sciences (CAS), Kunming 650216, P.R. China}
\affiliation{Key Laboratory for the Structure and Evolution of Celestial Objects, CAS, Kunming 650216, P.R. China}
\affiliation{Center for Astronomical Mega-Science, CAS, Beijing 100012, P.R. China}

\author{Zhanwen Han}
\affiliation{Yunnan Observatories, Chinese Academy of Sciences (CAS), Kunming 650216, P.R. China}
\affiliation{Key Laboratory for the Structure and Evolution of Celestial Objects, CAS, Kunming 650216, P.R. China}
\affiliation{University of Chinese Academy of Science, Beijing 100012, P.R. China}
\affiliation{Center for Astronomical Mega-Science, CAS, Beijing 100012, P.R. China}

\collaboration{4}{}
\begin{abstract}

Type Iax supernovae (SNe Iax) are an important sub-luminous class of SNe Ia. However, their progenitors and explosion mechanism have not been understood yet. It has been suggested that SNe Iax may be produced from weak deflagration explosions of Chandrasekhar-mass white dwarfs (WDs) in binary systems with a helium (He) star donor. In such progenitor systems, the stripped He companion material caused by the ejecta-companion interaction is expected to present some He features in their late-time spectra. However, the detection of He lines in late-time spectra of SNe Iax has not yet been successful, which gives an upper limit on the amount of stripped He mass of $\lesssim 2\times10^{-3}-0.1$\,\Msun. In this work, we study the interaction between SN Iax ejecta and a He star companion by performing three-dimensional hydrodynamical simulations with a weak pure deflagration explosion model. We find that about $\mathrm{4\times10^{-3}}$\,\Msun\ of He material can be stripped off from the companion star by SN explosion, which is very close to  (or lower than) the observational upper-limit on the total stripped He mass in SNe Iax. We, therefore, conclude that non-detection of He lines in late-time spectra of SNe Iax could be reasonably explained if they are indeed resulted from weak pure deflagration explosions of Chandrasekhar-mass WDs in progenitor systems with a He star donor.

\end{abstract}


\section{Introduction} \label{sec:intro}

Type Ia supernovae (SNe Ia) are generally thought to be the thermonuclear explosion of a white dwarf (WD) in a binary system \citep{Hoyle1960ApJ}. They play a crucial role in many areas of astrophysics. Using SNe Ia to measure cosmological distance and determine the cosmological parameters led to the discovery of the accelerating expansion of the Universe \citep{Riess1998AJ, Perlmutter1999ApJ}. But unfortunately, the nature of SN Ia progenitor systems and their explosion mechanism are still mysterious \citep[e.g., see][for a review]{Hillebrandt2000ARAA, Wang2012NAR, Maoz2014ARAA, Livio2018PhR}. Different progenitor models have been proposed for SNe Ia in the past few decades, comprising the single-degenerate (SD) model \citep{Whelan1973ApJ, Nomoto1982ApJb, Nomoto1984ApJ, Han2004MNRAS}, the double-degenerate (DD) model \citep{Iben1984ApJS, Webbink1984ApJ, Pakmor2010Nature}, and the sub-Chandrasekhar mass (sub-$M_\mathrm{ch}$) double-detonation model \citep{Nomoto1982ApJa, Woosley1986ApJ, Shen2007ApJ, Fink2007AAP, Sim2010ApJL, Gronow2020AAP} and so on.

Most SNe Ia follow the ``Phillips relation'' between decline rate and peak luminosity \citep{Phillips1993ApJ}, i.e., the so-called `normal' SNe Ia. However, many peculiar SNe Ia have been discovered to not follow the Phillips relation \citep{Li2003PASP}. SN~2002cx-like events, i.e., the so-called Type Iax supernovae (SNe Iax, \citealt{Foley2013ApJ}), are the most common peculiar SNe Ia, and they can contribute around $30\,\%$ of total SNe Ia \citep{Li2011Nature, Foley2013ApJ, White2015ApJ}. SNe Iax are much fainter than normal SNe Ia. They have a wide range of peak luminosities, $\mathrm{-14.2 \geqslant M_{V,peak} \gtrsim -18.5\,mag\,}$, and expansion velocities, $\mathrm{2000 \lesssim \vert v \vert \lesssim 8000\,km\,s^{-1}}$, near the peak luminosity (e.g., \citealt{Li2011Nature, Foley2013ApJ, Liu2015MNRAS, Jha2017Hsn}). They have shown the lack of a second maximum in the near-infrared light curves which is typically seen in normal SNe Ia \citep{Li2003PASP}. Typically, SNe Iax have a strong mixing ejecta \citep{Jha2006AJ}. In addition, two Iax events (i.e., SN~2004cs and SN~2007J) have been found to show helium (He) features, He~\textsc{i} emission, in their early-time spectra \citep{Rajala2005PASP, Foley2009AJ, Foley2013ApJ, Magee2019AAP}. Most SNe Iax are observed in late-type, star-forming galaxies, suggesting a short delay time of $\lesssim 100\,\mathrm{Myr}$ \citep{Foley2013ApJ, Lyman2013MNRAS, White2015ApJ}. However, one Iax event, SN 2008ge, was observed in an S0 galaxy with no signs of star formation \citep{Foley2010AJ, Foley2013ApJ}.

Many different explosion models have been proposed to explain SNe Iax \citep[e.g.,][]{Hoeflich1995ApJ, Moriya2010ApJ, Stritzinger2015AAP, Kromer2013MNRAS, Fink2014MNRAS}. Lower luminosities, explosion velocities and strong mixing of ejecta found in SNe Iax seem to suggest that SNe Iax may be produced from a deflagration explosion \citep{Branch2004PASP}. Recently, \citet{Jordan2012ApJL} investigated a weak deflagration explosion of a \Mch WD (the so-called ``failed-detonation model'' in their paper), in which the WD is partly burnt, leaving a bound remnant \citep[see also][]{Fink2014MNRAS}. They suggested that the low ejecta mass and velocity predicted by this model seem to be consistent with the observations of SNe Iax. Moreover, it has been shown that weak deflagration explosion of a \Mch CO WD and/or a hybrid CONe WD can well reproduce the observational features of Iax events, SN~2005hk (a proto-typical SN Iax) and SN~2008ha (the faint Iax event), respectively \citep{Kromer2013MNRAS, Kromer2015MNRAS}. \citet{Bulla2020AAP} also suggested that the maximum-light polarization signal seen in SN~2005hk can be well interpreted by asymmetries caused by SN explosion itself and by the ejecta-companion interaction in the weak deflagration explosion model. Therefore, the weak deflagration explosion of a \Mch WD seems to be a potential model for SNe Iax \citep[but see also][]{Hoeflich1995ApJ, Stritzinger2015AAP}.

The He lines seen in the spectra of SN~2004cs and SN~2007J indicate that their progenitor systems might contain some He material. \citet{McCully2014Nature} detected a bright source in the pre-explosion images of SN~2012Z. This bright source has been further suggested to be the He companion star of its progenitor system \citep{McCully2014Nature, Liu2015ApJ}. More interestingly, a bright source is still there four years after the explosion when the SN fades away, and it is brighter than that in pre-explosion images \citep{McCullyInprep}. In addition, SNe Iax generally have a short delay time of $\lesssim 100\,\mathrm{Myr}$ \citep{Lyman2013MNRAS, Lyman2018MNRAS}, which is consistent with the prediction of the WD~+~He star progenitor system \citep[e.g.,][]{Liu2015AAP}. Taking all this into account, the theoretical picture of an SD binary progenitor with a He companion star as the origin of SNe Iax seems to be consistent with observations.

In an SD progenitor system, the SN ejecta is expected to significantly interact with its companion star, stripping off some hydrogen (H)/He-rich material from the companion surface. The companion star will survive the explosion \citep{Wheeler1975ApJ}. If the stripped off H/He companion masses are massive enough, it would be expected to see some H/He features in late-time spectra of SNe Ia \citep[e.g.,][]{Leonard2007ApJ, Lundqvist2013MNRAS, Botyanszki2018ApJ}. The interactions of SN ejecta with a stellar companion star in SNe Ia have been investigated in detail by many two-dimensional/three-dimensional (2D/3D) hydrodynamical simulations \citep[e.g.,][]{Fryxell1981ApJ, Taam1984ApJ, Livne1992ApJ, Marietta2000ApJS, Pakmor2008AAP, Pan2010ApJ, Pan2012ApJ, Liu2012AAP, Liu2013AAP, Liu2013ApJa, Liu2013ApJb, Kutsuna2015PASJ, Boehner2017MNRAS, Bauer2019ApJ}. In particular, by adopting a weak pure deflagration explosion model of a \Mch WD produced by \citet{Kromer2013MNRAS}, the interaction of SN Iax ejecta with a main-sequence (MS) companion star has been addressed by \citet{Liu2013ApJb}. They found that a small amount of H-rich material can be stripped off from the MS companion star by the SN Iax explosion, which is consistent with the absence of H lines in the late-time spectra of SNe Iax.

Very recently, many observations have been carried out to search for the late-time He lines caused by the stripped He masses in the WD~+~He star progenitor systems. However, no He line has been detected yet \citep{Foley2016MNRAS, Jacobson2019MNRAS, Magee2019AAP, Tucker2019MNRAS}, placing an upper limit on the stripped He masses of $\lesssim 2\times10^{-3}-0.1$\,\Msun. This poses a serious challenge to the WD~+~He star as progenitor systems of SNe Iax. However, no theoretical modeling for the predictions of the stripped companion masses for the He star donor SD progenitor system has been done yet. It is still unclear whether the stripped off He masses by SN Iax explosion are massive enough to cause the presence of He lines in SN Iax late-time spectra.

In this work, we perform 3D hydrodynamical simulations of the SN~ejecta-companion interaction for the WD~+~He star progenitor system by assuming that SNe Iax are produced from weak pure deflagration explosions of \Mch WDs as proposed by \citet{Kromer2013MNRAS}. The paper is structured as follows. In Section~\ref{sec:method and models}, we introduce the methods and models used in our simulations. The results are presented in Section~\ref{sec:res}. In Section~\ref{sec:discussion}, we discuss the effect of binary separation on the results and compare the results with the observations. Moreover, the surviving companion star, different explosion models and uncertainties are discussed. The summary and conclusion are given in Section~\ref{sec:summary}.

\section{Method and Model} \label{sec:method and models}

To carry out 3D hydrodynamical simulations of the interaction of SN ejecta with a He companion star, we use a 3D Smoothed Particle Hydrodynamics (SPH) \citep{Gingold1977MNRAS, Lucy1977AJ} code, \textsc{Stellar GADGET} \citep{Pakmor2012MNRAS}. The \textsc{Stellar GADGET} is a modified version of the \textsc{GADGET} code \citep{Springel2001NA, Springel2005MNRAS}. It has been successfully used for modeling the merger of two white dwarfs for SNe Ia and the SN Ia ejecta-companion interaction \citep[e.g.,][]{Pakmor2010Nature, Pakmor2012MNRAS, Liu2012AAP, Liu2013AAP}. 

\subsection{The explosion model}

In this work, we assume that SNe Iax are generally produced from a pure weak deflagration explosion of a \Mch WD as described by \citet{Fink2014MNRAS}. In particular, the so-called N5def model is used in our impact simulations \citep{Kromer2013MNRAS}. In the N5def model, the weak deflagration explosion cannot completely disintegrate the entire WD. Only a small amount of material of around $\mathrm{0.372}$\,\Msun\ is ejected, leaving a bound remnant WD of about $\mathrm{1.03}$\,\Msun. The kinetic energy of ejecta material from this explosion is around $\mathrm{1.34\times{10^{50}}\,erg}$, and about $0.158$\,\Msun\ nickel is produced. Again, it has been shown that this model can reproduce the observational features of SN~2005hk \citep{Kromer2013MNRAS}. Note that the bound remnant WD is not included in our impact simulations of this work.

\subsection{A He star companion model}

In the present work, the ``He01 model'' of \citet{Liu2013ApJa} is directly used (see their Figs.~1 and 3) to represent a He~star companion model at the moment of the SN explosion, although they focused on the SN ejecta-companion interaction in normal SNe Ia. By using the Eggleton’s stellar evolution code \citep{Eggleton1971MNRAS, Eggleton1972MNRAS, Eggleton1973MNRAS}, they performed detailed binary evolution calculations for the WD~+~He star progenitor system to trace the whole process of a WD accumulating He-rich companion material until the WD increases its mass to approach the \Mch limit (i.e., the SN explosion, see \citealt{Liu2013ApJa}). The mass transfer occurs when a He star companion fills its Roche lobe. At the moment of the SN explosion, the He01 model has a mass of $M_{2}=1.24$\,\Msun\ and a radius of $R_{2}=\mathrm{1.91\times{10}^{10}\,cm}$. The He companion star is still filling its Roche lobe when the WD is assumed to explode as an SN Ia. The system has an orbital separation of $A=\mathrm{5\times{10}^{10}\,cm}$ at this moment \citep{Liu2013ApJa}.

In one-dimensional (1D) detailed binary evolution calculations for the SD progenitor systems of \citet{Liu2013ApJa}, the SN explosion is assumed to happen when the WD has a mass close to the \Mch limit. However, the accreting WDs were treated as a mass point in their 1D calculations. Because the explosion mechanism of a \Mch WD has not been clearly understood yet, the \Mch WD (i.e., mass point) could explode like a normal SN Ia or lead to the weak deflagration explosion for an SN Iax. In this work, we directly use the He star companion model, the ``He01 model'', adopted for normal SNe Ia by \citet{Liu2013ApJa} as an input of impact simulations for SNe Iax.    

\subsection{Initial setup}

\begin{figure*}
	\centering
	\includegraphics[width=0.32\textwidth]{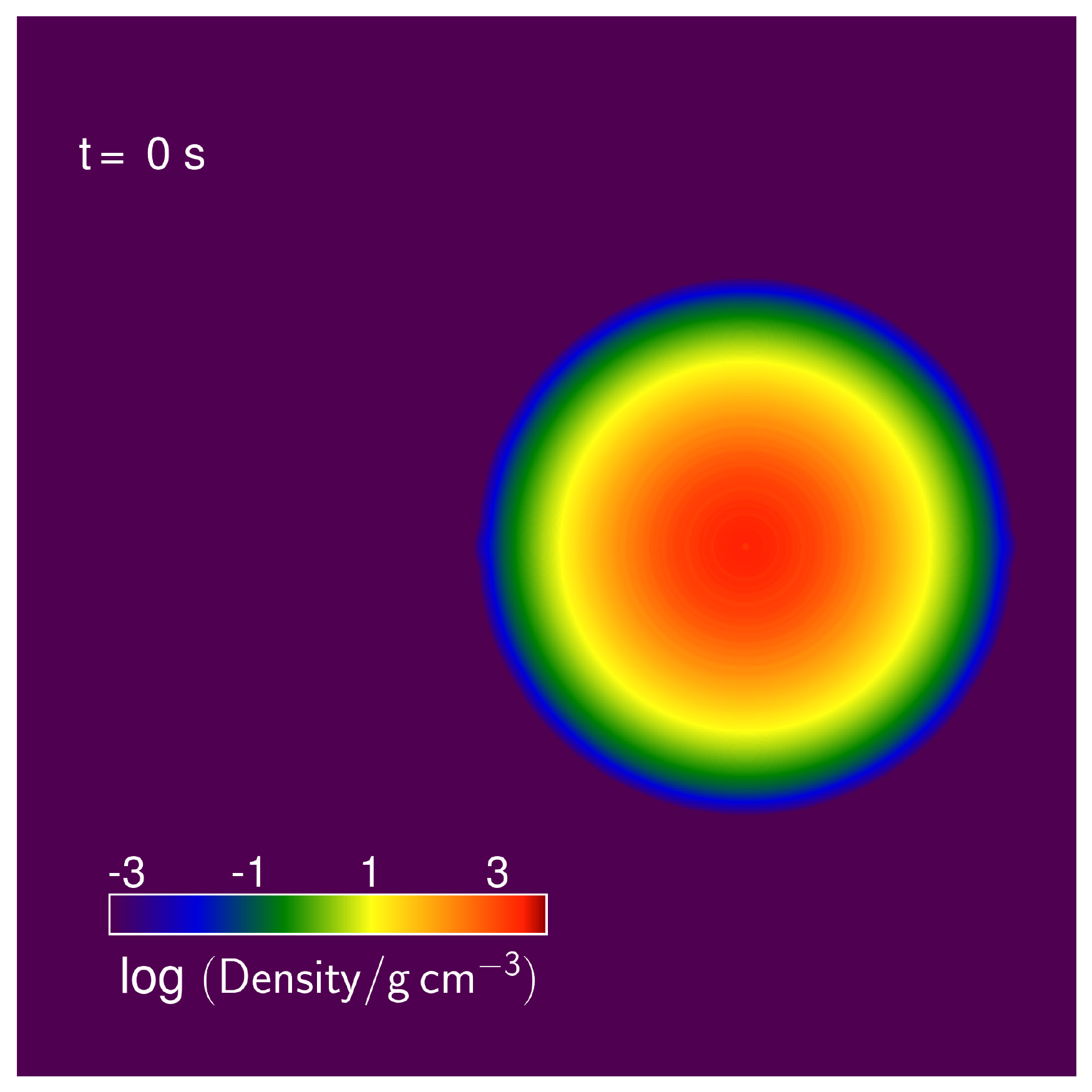}
	\includegraphics[width=0.32\textwidth]{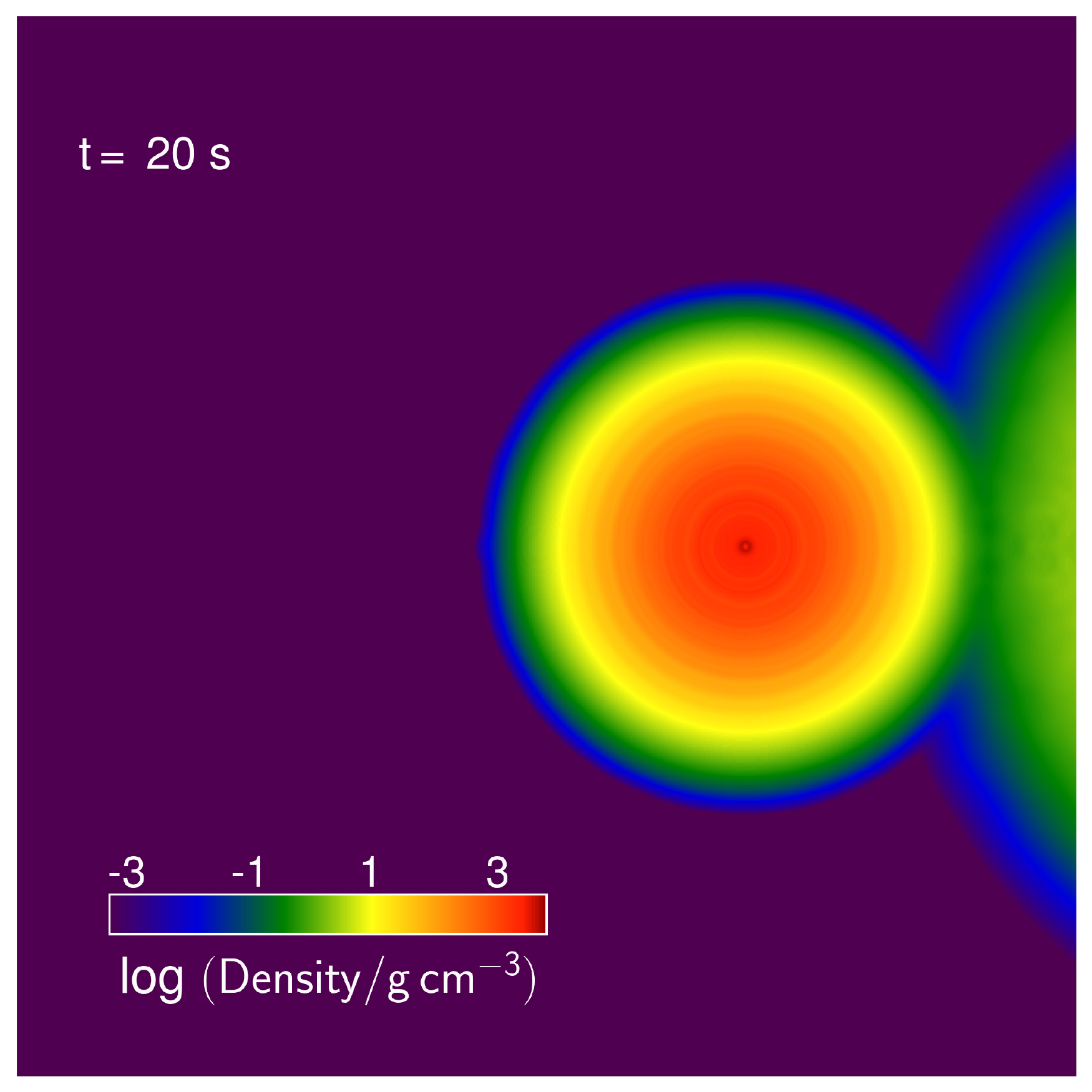}
	\includegraphics[width=0.32\textwidth]{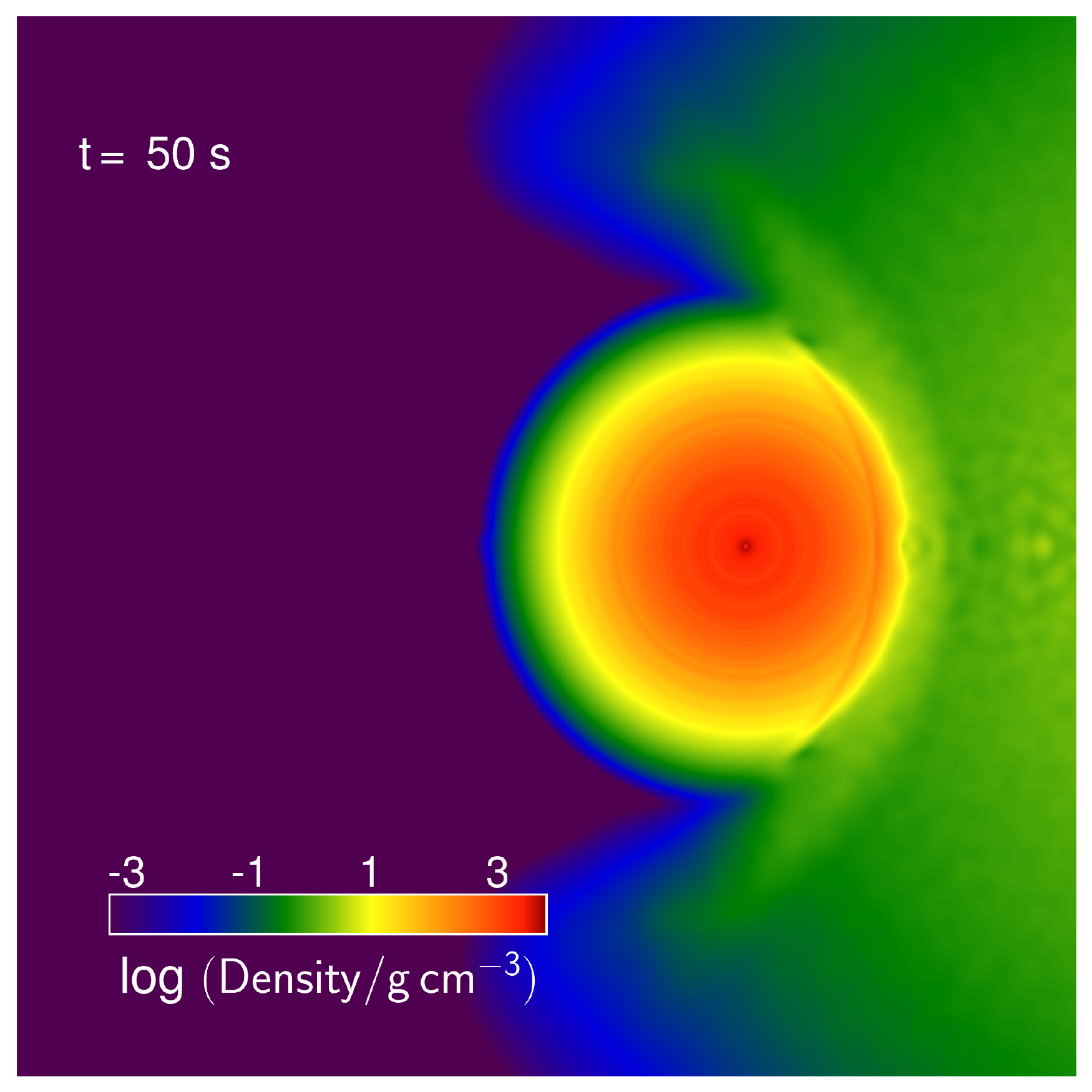}
	\includegraphics[width=0.32\textwidth]{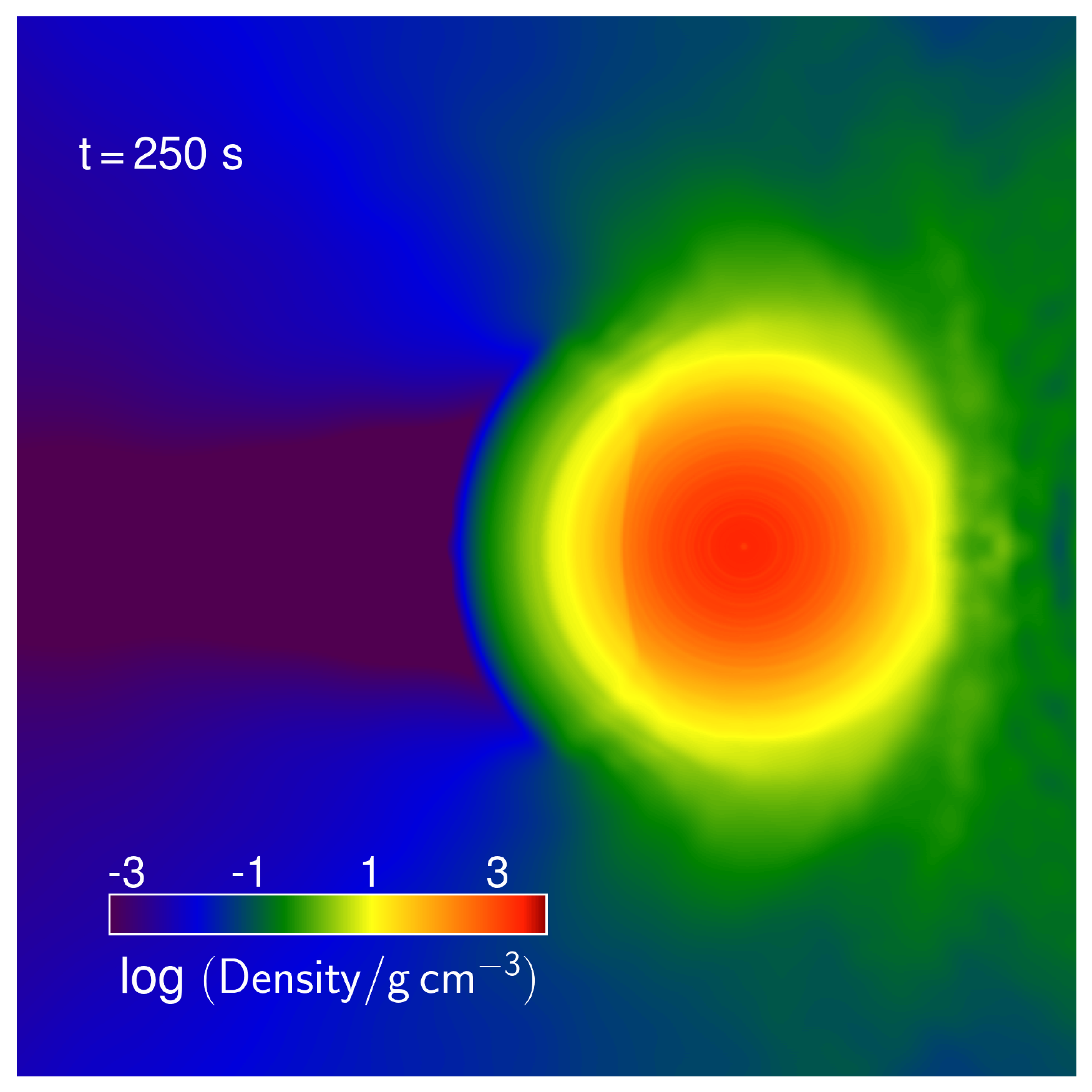}
	\includegraphics[width=0.32\textwidth]{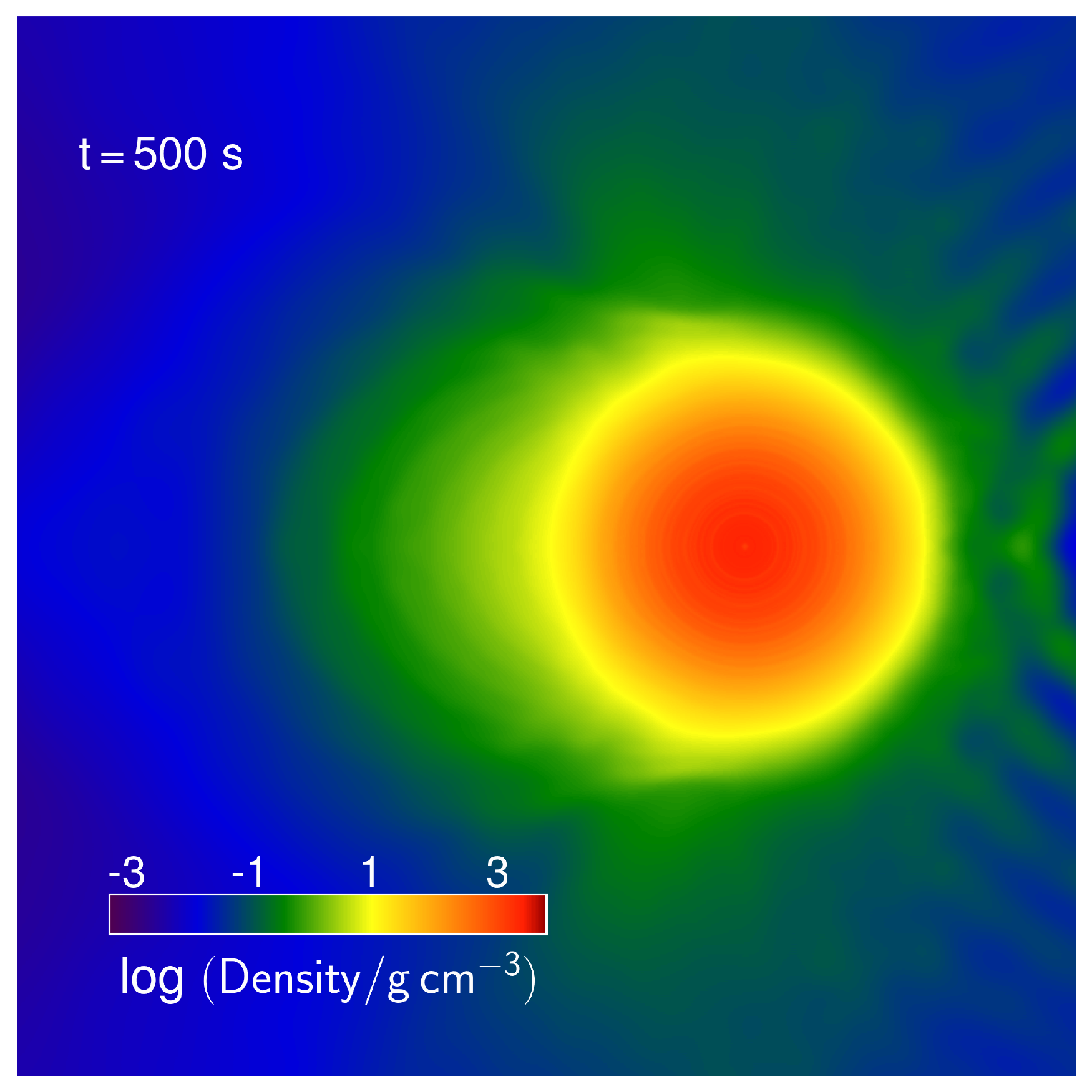}
	\includegraphics[width=0.32\textwidth]{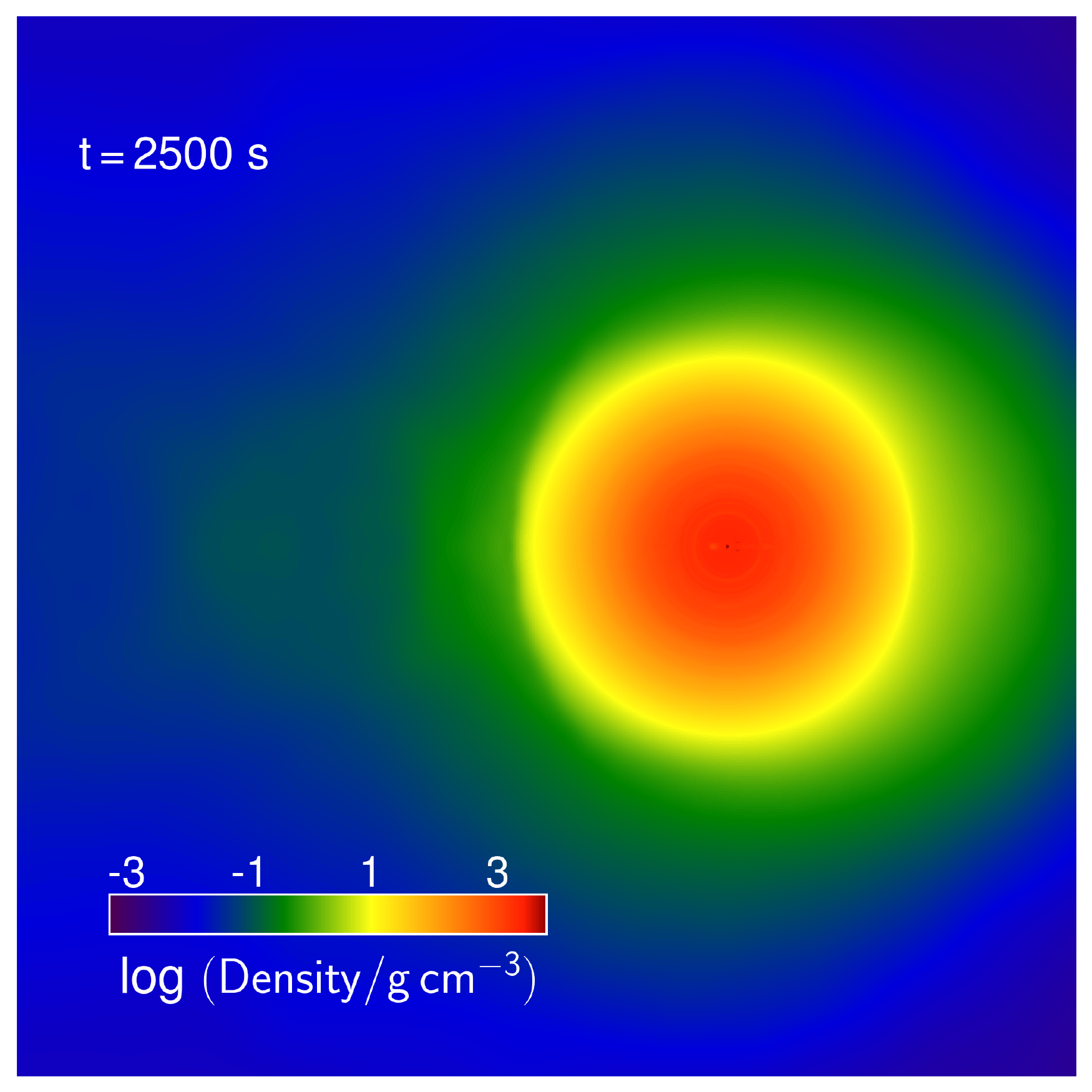}
	\caption{Slice-density distributions at $\mathrm{t=0\,s}$, $\mathrm{20\,s}$, $\mathrm{50\,s}$, $\mathrm{250\,s}$, $\mathrm{500\,s}$, and $\mathrm{2500\,s}$ after the N5def model is initiated in our standard impact simulation with a binary separation of $A=\mathrm{5.16\times{10}^{10}\,cm}$. The incoming SN ejecta is from $right$ to $left$. The color scale shows the logarithm of the mass density. \label{fig:1}}
\end{figure*}

All initial conditions and setup used in this work are the same as those in \citet{Liu2013ApJb}. The \textsc{Healpix} method \citep{Gorski2005ApJ} is used to map the 1D companion model (i.e., density and internal energy profile) to a particle distribution for our 3D SPH impact simulations \citep[see details in][]{Pakmor2012MNRAS}. In this work, we use 5 million particles to represent the He companion star in all simulations. All particles are set to have the same mass. It has been shown that using 5 million particles to set up the He companion star should be sufficient for the simulations of ejecta-companion interaction to study the amount of stripped companion mass \citep{Liu2013ApJb}. Because some numerical noise is introduced by initial mapping, we relax the initial 3D SPH companion model for several dynamical timescales to reduce such noise before we start the actual impact simulations.

Once the relaxation of our 3D SPH companion star is finished, the N5def model described above is added into the SPH code to represent an SN Iax explosion with a separation given by 1D binary evolution calculation (i.e., $A=\mathrm{5\times{10}^{10}\,cm}$). As all particles have the same mass, the total number of particles used for the N5def model is fixed. All impact simulations in this work are then simulated until the results such as the amount of stripped companion mass and kick velocity due to the ejecta-companion interaction reach a stable value as time goes. Therefore, all simulations ended at $\mathrm{5000\,s}$ after the SN explosion.

\section{Results} \label{sec:res}

In this section, we present the results of SN ejecta-companion interaction in our simulation with a binary separation of  $A=\mathrm{5.16\times{10}^{10}\,cm}$ ($M_{2}=1.24$\,\Msun, $R_{2}=\mathrm{1.91\times{10}^{10}\,cm}$). This simulation is defined as the ``standard simulation'' (i.e., Model4 in Table~\ref{tab:results}) in this work. By artificially changing the values of $A$ for a given explosion and companion model, we will investigate the dependency of numerical results on binary separations based on this standard simulation In section~\ref{sec:separation}.

\subsection{A description for SN ejecta-companion interaction}
\label{sub:description}

Figure~\ref{fig:1} shows the density distributions of all material at different times in our standard simulation\footnote{The data of this work will soon be publicly available online so that they can be accessible to the community to use. The data can also be obtained by directly contacting the relevant author.}. At the beginning of the simulation ($t=0\,\mathrm{s}$), the companion star is in an equilibrium state. The SN explodes on the right side of the companion star. At $t=20\,\mathrm{s}$, SN ejecta reaches the surface of the companion star. As SN ejecta collides with the star, a shock is being driven into the companion envelope and a bow shock starts to develop (at $t=50\,\mathrm{s}$), stripping some companion material from the side that is facing the SN explosion. As time goes by, the shock front starts to pass through the companion star. Because of the density gradient of the companion star, the shock center is decelerated due to the rising density as it moves towards the stellar core. At $t=500\,\mathrm{s}$, the shock has passed through the entire companion star, stripping more He material from the back of the companion star. The star is out of the hydrostatic equilibrium. At the same time, the star puffs up because of the significant shock heating during the interaction. About $t=2000\,\mathrm{s}$ after the explosion, the mass stripping is finished. The amount of stripped companion mass reaches the maximum and stays a constant, and the star starts to be back to the hydrostatic equilibrium state. At this point, the ejecta-companion interaction is finished.

\begin{figure}
		\centering
		\includegraphics[width=0.5\textwidth]{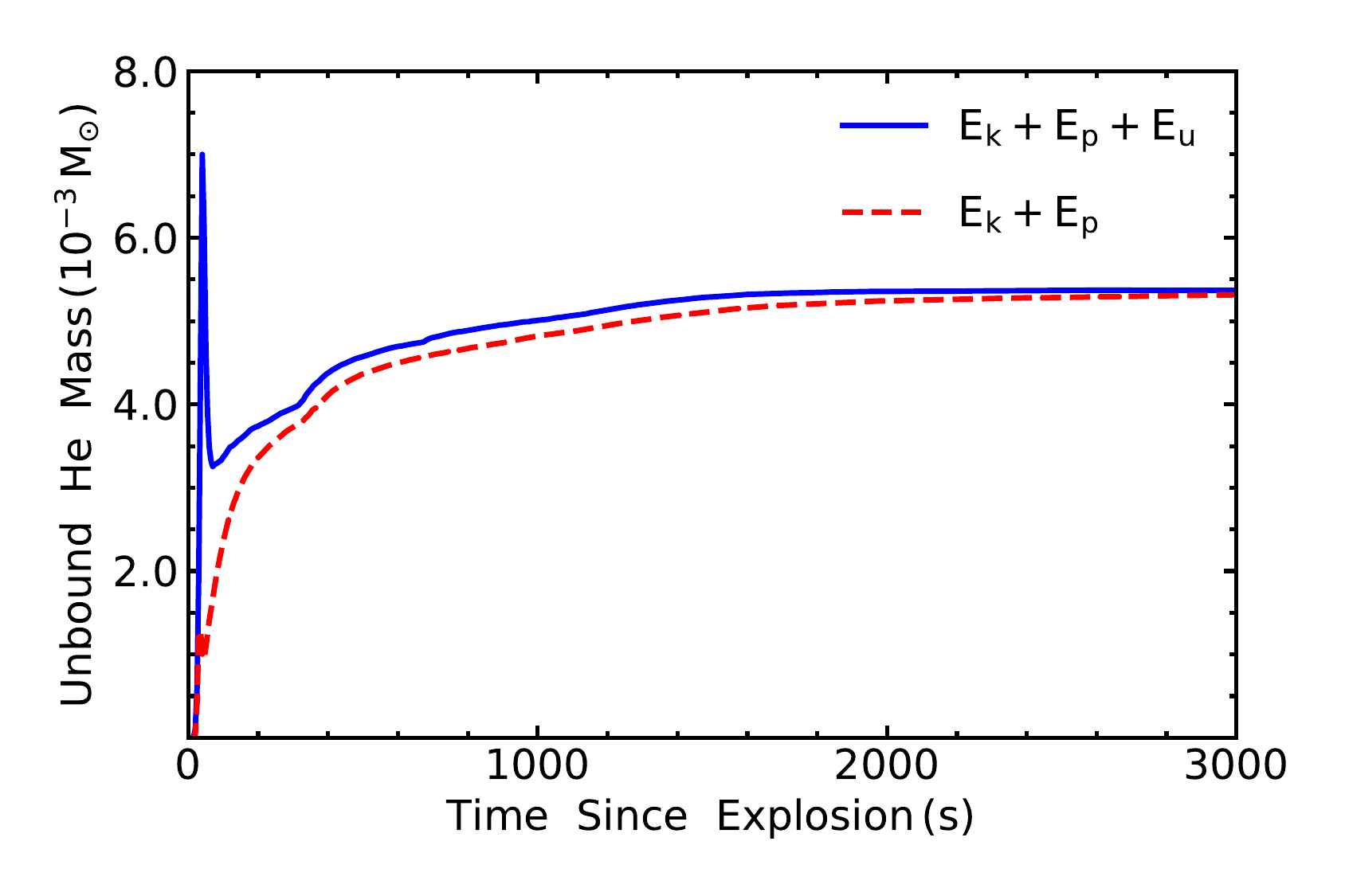}
		\caption{Amount of unbound companion mass caused by the SN explosion as a function of time. The results with (blue solid line) or without (red dashed line) including the internal energy (${E}_{\mathrm{u}}$) for calculating the total energy of each particle are shown for comparison.}
\label{fig:2}
\end{figure}

\begin{figure}
	\centering
	\includegraphics[width=0.5\textwidth]{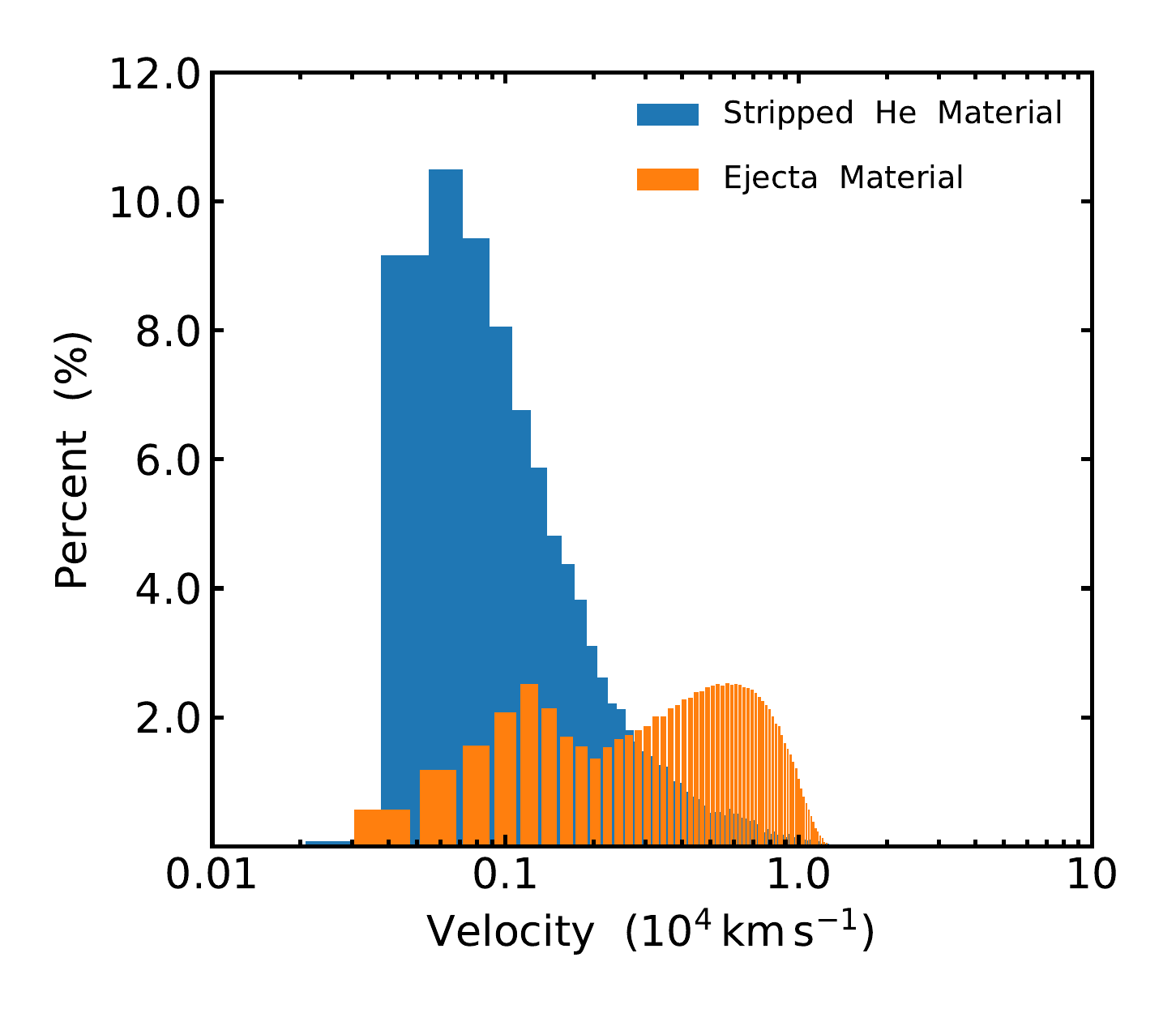}
	\caption{Velocity distribution of the He-rich companion material (blue) and SN ejecta (orange) at $\mathrm{5000\,s}$ in our simulations.}
\label{fig:3}
\end{figure}

\begin{figure}
	\centering	
			\includegraphics[width=0.5\textwidth]{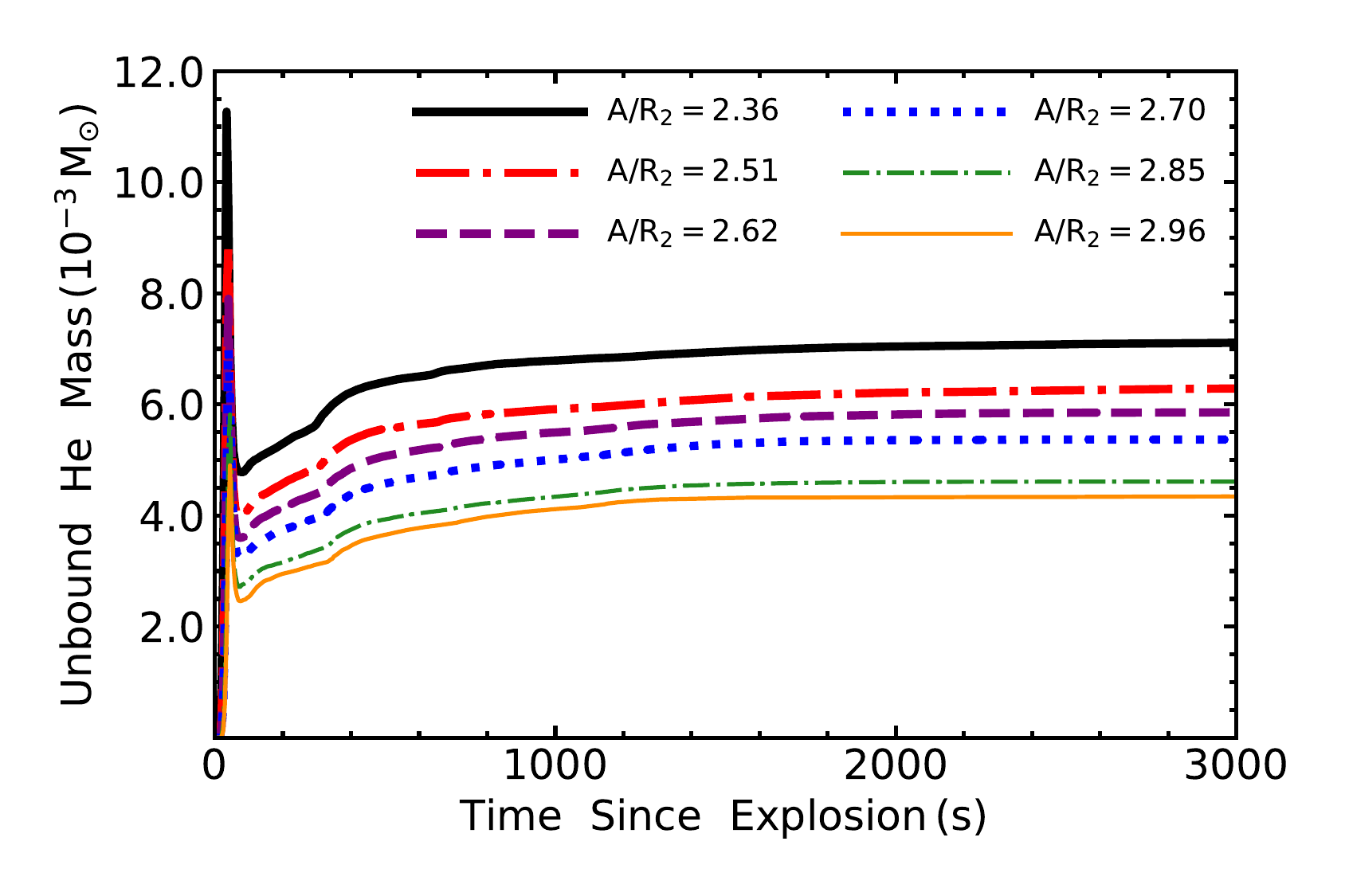}
		\caption{Unbound companion mass as a function of time for the simulations with different orbital separations.}
\label{fig:Hediff unbound mass}
\end{figure}

\setlength{\tabcolsep}{12pt}{
	\begin{deluxetable}{ccccc}
		\centering
		\caption{Results of our 3D impact simulations for different binary separations. \label{tab:results}}
		\tablehead{
			\colhead{$\mathrm{Model\  Name}$} & \colhead{$\mathrm{{A}/{R_2}}$} & \colhead{$M\mathrm{_{str}^{He}}$} & \colhead{ $V_\mathrm{{kick}}$} \\ 
			\colhead{} & \colhead{} & \colhead{($\mathrm{10^{-3}}$\,\Msun)} & \colhead{$\mathrm{(km\,s^{-1})}$}
		}
		\startdata
		$\mathrm{Model1}$ & $2.36$ & $7.16$ & $12.81$ \\
		$\mathrm{Model2}$ & $2.51$ & $6.31$ & $10.39$ \\
		$\mathrm{Model3}$ & $2.62$ & $5.87$ & $8.85$ \\
		$\mathrm{Model4}$ & $2.70$ & $5.38$ & $7.63$ \\
		$\mathrm{Model5}$ & $2.85$ & $4.62$ & $5.83$ \\
		$\mathrm{Model6}$ & $2.96$ & $4.35$ & $4.64$ \\
		\enddata
		\tablecomments{Here, $\mathit{{A}/{R_2}}$ is the ratio of binary separation and radius of the He companion star. $M_{\mathrm{str}}^{\mathrm{He}}$ is the amount of stripped companion mass, and $V_\mathrm{{kick}}$ is the kick velocity of the companion star. }
	\end{deluxetable}
}

\subsection{Stripped He mass} \label{subsec:stripped mass}

The amount of stripped companion mass is calculated by summing all unbound companion particles which have a positive total energy. The total energy of each particle is calculated by summing its kinetic energy (${E}_\mathrm{k}>0$), gravitational potential energy (${E}_\mathrm{p}<0$) and internal energy (${E}_\mathrm{u}>0$).

\begin{figure*}
		\centering
		\includegraphics[width=0.48\textwidth]{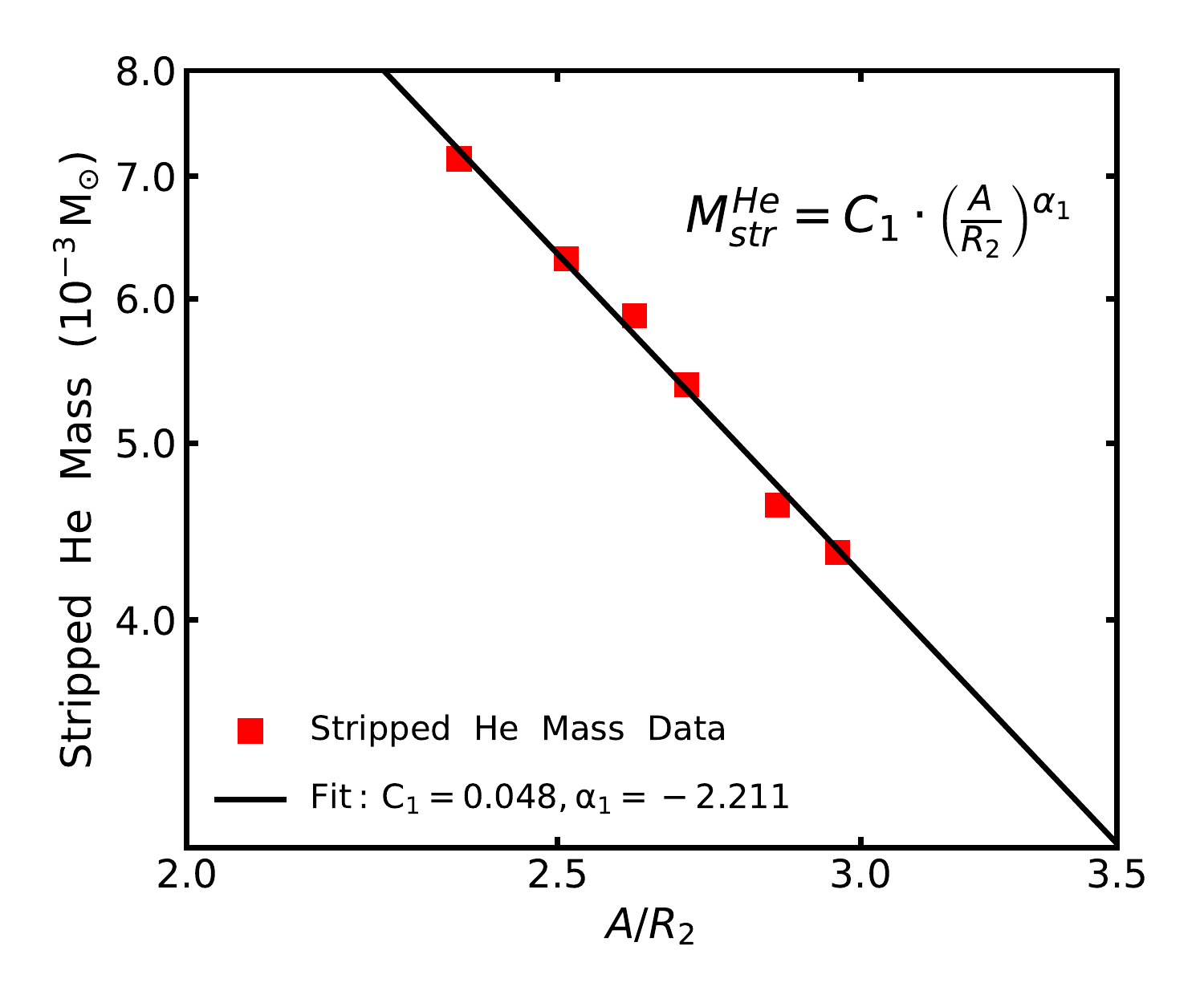}
		\includegraphics[width=0.48\textwidth]{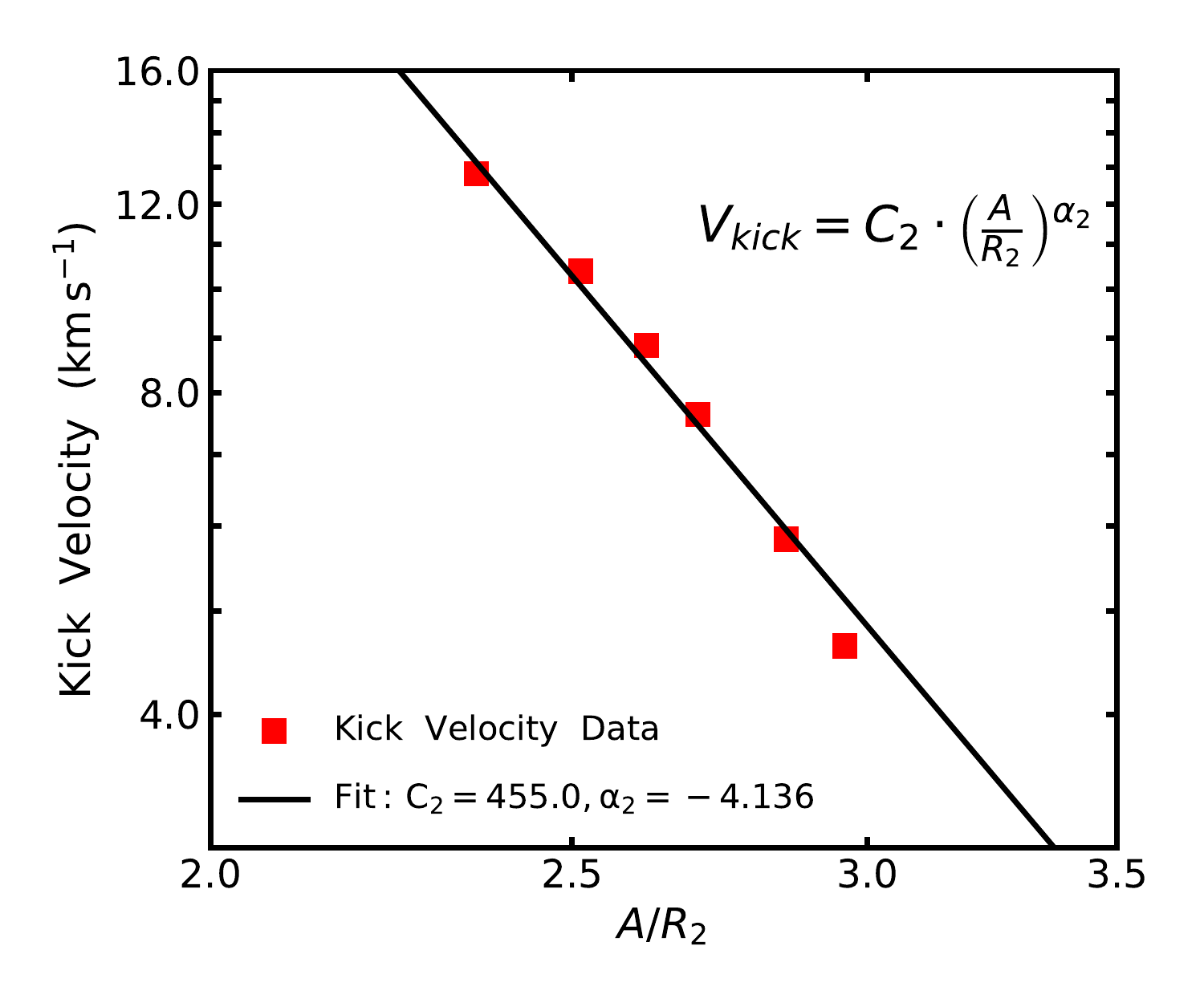}
		\caption{\textit{Left panel:} The relation between the total stripped companion mass ($M_{\mathrm{str}}^{\mathrm{He}}$) and the ratio of binary separation to the companion radius (${{A}/{R_2}}$). \textit{Right panel:} similar to the \textit{left panel}, but for the kick velocity, ${V_{\mathrm{kick}}}$. Both relations can be fitted by a power-law. Fitting parameters are given in figures.}
		\label{fig:5}
\end{figure*}

Figure~\ref{fig:2} shows the amount of stripped He mass from the surface of the companion star by SN explosion as a function of time for our standard simulation (i.e., $A/R_{2}=2.62$). As it is shown, the amount of stripped companion mass reaches a stable value of $\mathrm{5.4\times10^{-3}}$\,\Msun\ about $3000$ seconds after the explosion. This means that the ejecta-companion interaction is finished at this moment. As a result, we obtain that the total amount of about $\mathrm{5.4\times10^{-3}}$\,\Msun\ He-rich material can be stripped off from the companion surface.

Figure~\ref{fig:3} illustrates the velocity distributions of the stripped companion material (blue region) and SN ejecta (orange region) at the end of our simulation. It is shown that the stripped companion material has a typical velocity of $\sim 600-700\,\mathrm{km\,s^{-1}}$, which is slower than that of post-interaction SN ejecta ($\sim 7000\,\mathrm{km\,s^{-1}}$) by a factor of about ten. One can predict that most of the stripped companion material will be surrounded by SN ejecta as time goes by. It means that the He lines caused by the stripped He material can only be visible at a late-time phase as the photosphere moves inward from outside of ejecta to reach the region of stripped material ($\gtrsim100\,\mathrm{days}$ after the explosion).

\section{Discussion} \label{sec:discussion}

\subsection{Binary separation dependency}
\label{sec:separation}

For a given explosion and companion star model, it has been shown that the ratio of binary separation to companion radius, $A/R_{2}$, is the key factor in determining the results of the SN ejecta-companion interaction \citep[e.g.,][]{Marietta2000ApJS, Pakmor2008AAP, Liu2012AAP}. Binary population synthesis (BPS) calculations for the WD~+~He star progenitor scenario of SNe Ia have obtained that the value of $A/R_{2}$ at the time of SN explosion covers a range of $\sim2.5-3.0$ \citep[e.g., see Fig.~8 of][]{Liu2013ApJa}. Our standard simulation has a value of $A/R_{2}=2.62$. To investigate the effect of different binary separations on the results in our impact simulations, we artificially change the separation ($A$) between SN and He companion star based on our typical simulations to cover a range of $A/R_{2}$ from 2.36 to 2.96 (see Table~\ref{tab:results}). This means that all other parameters are kept the same expect for the values of $A$.

Figure~\ref{fig:Hediff unbound mass} presents the stripped companion masses as a function of time in our simulations for different orbital separations. Also, the total amount of stripped companion masses ($M\mathrm{_{str}^{He}}$) and the kick velocity received by the companion star because of SN impact ($V_\mathrm{kick}$) at the end of the simulations are given in Table~\ref{tab:results}. We have furthermore explored the relationship between the amount of stripped masses (and/or kick velocity) and orbital separation, finding that they can be well fitted by a power low (see Fig.~\ref{fig:5}). This is consistent with the results of previous studies \citep[e.g.,][]{Marietta2000ApJS, Pakmor2008AAP, Liu2012AAP, Pan2012ApJ}.

In BPS calculations for the WD~+~He star progenitor scenario, the typical value of $A/R_{2}$ at the moment of SN explosion is about $3.0$ \citep[e.g., see Fig.~8 of][]{Liu2013ApJa}. By using the fitting functions given in Fig.~\ref{fig:5}, we obtain that the typical amount of stripped He masses for SNe Iax in our studied channel of this work should be about $4\times10^{-3}$\,\Msun\ (see also in Table~\ref{tab:results}). Again, current observations have not detected the He lines in late-time spectra of SNe Iax yet, which gives an upper limit on the stripped He masses of $\lesssim 2\times10^{-3}-0.1$\,\Msun\ \citep{Foley2016MNRAS, Jacobson2019MNRAS, Magee2019AAP, Tucker2019MNRAS}. The typical amount of stripped He masses ($4\times10^{-3}$\,\Msun) predicted in our simulations is very close to (or lower than) the observational upper limit. Therefore, we conclude that the non-detection of He lines in late-time spectra of SNe Iax is because a small amount of stripped He masses is caused by SN explosion. This supports that SNe Iax are produced from the binary progenitor system consists of a WD and a He star companion.

By studying the interaction of SN ejecta with a He companion star for normal SNe Ia, \citet{Liu2013ApJa} have found that  about $>1.9\times10^{-2}$\,\Msun\ He masses can be stripped off from the surface of a companion star (see also \citealt{Pan2010ApJ, Pan2012ApJ}). Their stripped He masses are higher than ours by one order of magnitude. This is because they used the W7 model to represent SN explosion for normal events. However, we focus on the subluminous SNe Iax and the weak deflagration explosion is assumed in our work. The W7 model has total explosion energy of $1.23\times10^{51}\,\mathrm{erg}$, which is higher than that of the N5def model of $1.34\times10^{50}\,\mathrm{erg}$ by one order of magnitude, leading to that more companion material has been stripped off by SN explosion in their simulations. Note that the amount of stripped companion masses has been found to linearly increase as the explosion energy increases \citep{Pakmor2008AAP, Liu2012AAP, Liu2013ApJb}.

\subsection{Surviving companion stars}
   
If SNe Iax were indeed produced from weak deflagration explosions of \Mch WDs in SD progenitor systems with a He star companion, the He companion stars would be expected to survive from the explosion and show some special observational features such as high spatial velocities and the enrichment of heavy elements. Therefore, searching for the surviving He companion star provides a way to place constraints on this progenitor model. Again, \citet{McCully2014Nature} detected a blue luminous source in the pre-explosion image of SN~2012Z. This has been further suggested to be consistent with the predictions of a He star donor progenitor system. The follow-up observations have detected a brighter source there a few years ($\sim1500$ days) after the explosion. It is even brighter than the pre-explosion detection and a normal Ia event, SN~2011fe, at that phase \citep{McCullyInprep}.

The surviving companion star in the SD progenitor system will be significantly shock-heated during the ejecta-companion interaction. This could lead to that the companion star puffs up dramatically and becomes overluminous after the explosion \citep{Pan2013ApJ, Shappee2013ApJ}. By tracing post-explosion evolution of a He companion, i.e., the so-called ``HeWDd'' model in their paper, \citet{Pan2013ApJ} found that the post-impact luminosity of this He star can reach $\sim\mathrm{10^{4}}$\,$\mathit{L_{\sun}}$ about $10\,\mathrm{yrs}$ after the explosion, which is brighter than its pre-SN luminosity ($\sim10\,\mathit{L_{\sun}}$) by a factor of about $10^{3}$. The He star companion model used in the present work is quite similar to the ``HeWDd'' model. However, the explosion model, i.e., the N5def model, used in our impact simulations has a lower explosion energy than that (i.e., the W7 model, see \citealt{Nomoto1984ApJ}) adopted by \citet{Pan2013ApJ}. This leads to that our post-impact He star has a less energy deposition by shock heating during the ejecta-companion interaction. Therefore, we can roughly predict that our surviving He companion star would be less luminous than the ``HeWDd'' model of \citet{Pan2013ApJ} in its post-explosion evolution. In a forthcoming study, we will use a 1D stellar evolution code to trace the long-term post-explosion evolution of surviving He stars from our 3D impact simulations to examine whether they could provide a good explanation for the post-SN brighter source detected in SN~2012Z.

\subsection{Different explosion models}

SNe Iax have a wide range of peak luminosities ($\mathrm{-14.2 \geqslant M_{V,peak} \gtrsim -18.5\,mag}$, e.g., \citealt{Li2011Nature, Foley2013ApJ, Jha2017Hsn}), including the brighter class members (e.g., SN~2012Z, $M_\mathrm{V,peak}= -18.5\,\mathrm{mag}$, see \citealt{Stritzinger2015AAP}) and the faintest event (i.e., SN~2008ha, $M_\mathrm{V,peak}= -14.2\,\mathrm{mag}$, see \citealt{Foley2009AJ, Foley2010AJ}). The N5def model used in this work has been shown to reproduce observational features of the typical SNe Iax such as SN~2005hk ($M_\mathrm{V,peak}= -18.1\,\mathrm{mag}$) well \citep{Kromer2013MNRAS}. However, to provide a reasonable agreement with the observables of the faintest event, SN~2008ha, an off-centre deflagration explosion in a near \Mch hybrid CONe WD is needed \citep{Kromer2015MNRAS}. In this deflagration model, only $0.014$\,\Msun\ of a \Mch hybrid CONe WD is ejected with asymptotic kinetic energy of $E_\mathrm{K}=1.8\times10^{48}\,\mathrm{erg}$ (which is about two orders of magnitude lower than that of the N5def model), leaving a massive bound remnant of $1.39$\,\Msun\ after the explosion \citep{Kromer2010ApJ}. If this specific deflagration model for SN~2008ha is used for our impact simulation, we would expect that the amount of stripped He masses decreases by about one/two orders of magnitude compared with that in this work because of its lower explosion energy \citep{Liu2012AAP, Liu2013AAP}. 

In addition, \citet{Stritzinger2015AAP} suggested that a near \Mch WD progenitor experiencing a pulsational delayed detonation (PDD, \citealt{Hoeflich1995ApJ}) seems to be also a good candidate model for SNe Iax such as SN~2012Z. The amount of stripped He masses could be different from the predictions in this work if the PDD model is used for our impact simulations, which needs to be investigated in detail by future studies.

\subsection{Uncertainties in our modelling}

In this work, the He companion star model is fixed in all simulations with different binary separations. However, it has been found that the detailed structures of the companion star (i.e., at different evolutionary phases when the SN explodes) can affect the amount of stripped companion material for a given explosion model and value of $A/R_{2}$ \citep[e.g.,][]{Liu2012AAP, Pan2012ApJ}. For instance, \citet{Liu2012AAP} have shown that the amount of stripped H mass from a MS companion model could be changed by a factor of 2 for a given $A/R_{2}$ if the companion star is slightly evolved (see their Fig.~6).

Moreover, the orbital motion and rotation are not included in our simulations, although we do not expect that both orbital motion and rotation will significantly affect the amount of stripped companion mass and the kick velocity because they are generally slower than the typical expansion velocity of SN ejecta of $7000\,\mathrm{km\,s^{-1}}$ by one order of magnitude \citep[e.g.,][]{Pan2012ApJ, Liu2013ApJa}. Future modeling still needs to be expanded (or improved) to cover a wider range of explosion and companion models for SNe Iax with including the orbital motion and rotation to make better predictions on the amount of stripped He mass by SN explosion.

\section{Summary and Conclusion} \label{sec:summary}

In this work, by assuming that SNe Iax are generally generated from weak deflagration explosions of \Mch WDs in the WD~+~He star progenitor systems \citep{Kromer2013MNRAS}, we have investigated the interaction of SN Iax ejecta with a He companion star by performing 3D hydrodynamical simulations with the \textsc{Stellar GADGET} code \citep{Springel2001NA, Pakmor2012MNRAS}. For the companion star model at the moment of the explosion, we directly use the one created by \citet{Liu2013ApJa} with the Eggleton’s stellar evolution code \citep{Eggleton1971MNRAS, Eggleton1972MNRAS, Eggleton1973MNRAS}. Our results and conclusions can be summarized as follows. 

\vspace{-\topsep}
\begin{itemize}

\item[(1)] For our standard simulation (i.e., the companion mass, the companion radius and the orbital separation are $M_{2}=1.24$\,\Msun, $R_{2}=\mathrm{1.91\times{10}^{10}\,cm}$, and $A=\mathrm{5.16\times{10}^{10}\,cm}$, respectively), it is found that about $5.4\times10^{-3}$\,\Msun\ He-rich material is stripped off from the He star companion. This corresponds to about $0.4$ percent of the initial mass of a He star.

\item[(2)] The stripped companion material moves with characteristic speeds of $600-700$\,$\mathrm{km\,s^{-1}}$, which is slower than the typical velocity of SN ejecta of $\mathrm{7000\,km\,s^{-1}}$ (see Fig.~\ref{fig:3}).

\item[(3)] It is found that the amount of stripped He mass ($M\mathrm{_{str}^{He}}$) and kick velocity ($V_\mathrm{kick}$) decrease as the binary orbital separation increases, which is in good agreement with the power-law relation (see Fig.~\ref{fig:5}). This is consistent with the results predicted by other groups for normal SNe Ia.

\item[(4)] Current BPS calculations predict that the parameter of $A/R_{2}$ at the time of SN explosion in the WD~+~He star progenitor scenario covers a wide range of $2.5-3.0$ and peaks at a typical value of $A/R_{2}=3.0$  (see Fig.~8 of \citealt{Liu2013ApJa}). According to the derived power-law relation between stripped mass and  $A/R_{2}$, we can further predict that the amount of stripped He mass in SNe Iax has a peak of $4\times10^{-3}$\,\Msun, which is very close to (or lower than) the observational upper-limit on the stripped He masses of $\lesssim 2\times10^{-3}-0.1$\,\Msun\ \citep[e.g.,][]{Foley2016MNRAS, Jacobson2019MNRAS, Magee2019AAP, Tucker2019MNRAS}. This suggests that the stripped He might be hidden in the late-time spectra of SNe Iax if we assume that SNe Iax are generally produced from weak deflagration explosions of \Mch WDs in the WD~+~He star progenitor systems.

\item[(5)] The simulations show that the SD He companion model is consistent with the observational upper limits of mass stripping. This suggests that the WD~+~He star system may be a potential progenitor of SNe Iax.


\end{itemize}

Our simulations do not cover different possible explosion scenarios proposed for SNe Iax. Future research will focus on investigating the SN ejecta-companion interaction in SNe Iax by adopting different explosion models. In addition, the long-term evolution of surviving companion stars from our hydrodynamical simulations will be studied in the forthcoming paper. This will be expected to provide a strict constraint on the amount of stripped He mass and to comprehensively make predictions on the observational features of surviving companion stars, and therefore help to examine the reliability of theoretical models for SNe Iax by observing their late-time spectra and the surviving companion star.

\acknowledgments

We thank the anonymous referee for the helpful comments. We would like to thank Jiangdan Li for helping to improve the language of this paper. We thank Callum McCutcheon, Xiangcun Meng and Hailiang Chen for their useful discussions. YTZ would like to thank Jiao Li for helping making the figures. This work made use of the Heidelberg Supernova Model Archive \citep[HESMA,][see \url{https://hesma.h-its.org}]{Kromer2017MmSAI}. The authors also gratefully acknowledge the part of computing time granted by the Yunnan Observatories and provided on the facilities at the Yunnan Observatories Supercomputing Platform. This work is supported by the National Natural Science Foundation of China (NSFC, Nos. 11873016, 11521303 and 11733008), the Chinese Academy of Sciences, and Yunnan Province (Nos. 2019HA012 and 2017HC018).\\
\software{yt-project \citep[version 3.4.1][]{Turk2011ApJS}, matplotlib \citep[version 3.1.0][]{Hunter2007CSE}, numpy \citep[version 1.16.4][]{Van2011CSE} and scipy \citep[version 1.1.0][]{Virtanen2019arXiv}}

\bibliography{ref}{}
\bibliographystyle{aasjournal}

\end{document}